\documentclass[
preprintnumbers,
preprint,
a4paper,
showpacs, 
amsmath, 
amssymb, 
floatfix,
nofootinbib,
]{revtex4}

\usepackage[dvips]{graphicx}

\begin{document}

\title{Solitonlike solutions of magnetostatic equilibria:\\
Plane-symmetric case}

\author{Hirotaka Yoshino}

\affiliation{Department of Physics, University of Alberta, 
Edmonton, Alberta, Canada T6G 2G7}

\author{Kohei Onda}

\affiliation{Department of Physics, Graduate School of Science, Nagoya
University, Chikusa, Nagoya 464-8602, Japan}

\preprint{Alberta-Thy-14-08}

\date{May 28, 2008}

%
%
\begin{abstract}
We present the plane-symmetric solitonlike solutions of 
magnetostatic equilibria by solving the nonlinear 
Grad-Shafranov (GS) equation numerically. 
The solutions have solitonlike and periodic structures in the $x$ 
and $y$ directions, respectively, and $z$ is the direction of plane symmetry. 
Although such solutions are unstable against the numerical iteration,
we give the procedure to realize the sufficient convergence.
Our result provides the definite answer for the existence
of the solitonlike solutions that was questioned in recent years.
The method developed in this paper will make it possible
to study the axisymmetric solitonlike solutions of the
nonlinear GS equation, which could model
astrophysical jets with knotty structures.
\end{abstract}

\maketitle

%
%
\section{Introduction}

The magnetostatic equilibria are of fundamental interest,
since they well approximate slowly varying magnetically
confined plasma configurations. 
In systems with the helical symmetry (i.e. unification
of plane symmetry and axisymmetry),
the magnetostatic equilibria are described by the
so-called Grad-Shafranov (GS) equation 
(see \cite{Biskamp93} for a review). 
The GS equation is an elliptic equation for the flux function $\Psi$ 
with a source term depending on two functions of $\Psi$
that can be chosen freely. The solutions of 
the GS equation are often used in theoretical studies
in the contexts of the astrophysics or the tokamak physics.

Recently, the existence of an interesting solution
of the GS equation was suggested by Lapenta \cite{Lap03}.
In that paper, the GS equation with a cubic source term 
(say, the cubic GS equation)
in the plane-symmetric case was discussed. 
An analogy between this equation and the cubic Schr\"odinger
equation was pointed out, and the real part of the 
solution of the cubic Schr\"odinger equation was 
presented as an analytic ``solution'' to the GS equation.
This ``solution'' is periodic in the $y$ direction,
and has a solitonlike structure in the $x$ direction.
Here, $z$ is the direction of the plane-symmetry.
Unfortunately, an erroneous assumption in this analysis 
was pointed out \cite{THT04} and thus the 
``solution'' of \cite{Lap03} cannot be accepted. 
However, Lapenta performed a numerical simulation
adopting his ``solution'' as the initial state 
and observed that the system is relaxed to a quasi-equilibrium state 
which maintains the solitonlike structure \cite{Lap04} 
(until the instability becomes relevant). 
Then, he claimed that the solitonlike solution has practical
applicability and relevance. The discussion that supports the existence
of the solitonlike solution was given in \cite{SSP05}.
The growth of the instability of the solitonlike structure
of this system was simulated in order to discuss the collimation
and expansion of astrophysical jets \cite{LK05}.

In this paper, we present the plane-symmetric 
solitonlike solutions of the cubic GS equation 
by performing highly accurate numerical calculations. 
There are several reasons for doing so. 
First, the question on the existence 
of the solitonlike solutions of the GS equation should be answered.
Although the simulations in \cite{Lap04, LK05}
strongly indicate the existence of the solitonlike solutions 
and explain the gross features of such solutions,
strictly speaking, 
the obtained quasi-equilibrium states are not
the solutions of the GS equation since they 
weakly depend on time. Besides, 
even if we ignore the time dependence, 
the GS equations which the obtained 
quasi-equilibrium states satisfy are not necessarily 
the cubic GS equation, i.e., the form of the source
term should be different in general. 
Since the possible existence of the solitonlike solutions 
attracts attentions, it is worth studying these solutions
by directly solving the cubic GS equation.

Second, the numerical technique for solving the
nonlinear GS equation should be developed. 
As we will see later, there is a difficulty 
in solving the cubic GS equation 
such that the nontrivial solution of this equation is 
unstable against the numerical iteration, 
i.e. the standard technique to solve 
elliptic partial differential equations. 
The development of such a technique is important
not only in the plane-symmetric case
but also in the axisymmetric case,
because the solitonlike solution of the axisymmetric 
GS equation is expected to have interesting astrophysical applications.
In fact, the observations of active galactic nuclei (AGN) suggest 
that the astrophysical jets have the magnetic multiple islands \cite{BHLBL94}.
Several models for such knotty jets have been proposed,
and one possible direction is to model the knotty jets
as the magnetostatic equilibria in the comoving frames 
\cite{Lap03, Lap04, LK05, KC85, B00}.
As the first step to study the axisymmetric nonlinear GS equation, 
it is useful to begin with the simpler plane-symmetric case
where the existence of solitonlike solutions is highly likely.
In this paper, we give a procedure to realize the sufficient convergence 
and successfully obtain the numerically unstable solutions.
Here, it has to be mentioned that some isolated axisymmetric 
toroidal Alfv\'en solitons were numerically obtained
in \cite{PPC82, PPS86} by combining
Fourier transformation and the method of the Green's function.
Although their method can be applicable
also for the present cases, our method is somewhat
simpler and easier.

This paper is organized as follows. 
In the next section, we briefly review the GS equation 
and introduce the cubic GS equation. The asymptotic
behavior of the solitonlike solution is also studied.
In Sec. III, we explain the numerical method and estimate
the numerical errors. 
In Sec. IV, the numerical results are presented.
Some properties of the obtained solution are also examined.
Sec. V is devoted to summary and discussion.
In this paper, we adopt the unit where the vacuum permeability 
$\mu=1$ because it can be restored by dimensional considerations 
if necessary.

%
%
\section{The Grad-Shafranov equation}

In this section, we review the Grad-Shafranov (GS) equation
for the plane-symmetric case and introduce the
assumption on the arbitrary functions that leads to the cubic GS equation. 
Then we explain our requirements on the behavior of the solution.

We consider static configurations where the magnetic
fields $\mathbf{B}$ are embedded in 
an ideal plasma with velocity $\mathbf{v}=0$.
The basic equations are Ampere's law and the
force balance equation:
%
\begin{equation}
\mathbf{J}=\nabla\times\mathbf{B},
\label{Ampere}
\end{equation}
%
%
\begin{equation}
\mathbf{J}\times\mathbf{B}=\nabla p,
\label{Force-balance}
\end{equation}
%
with the Gauss law $\nabla\cdot\mathbf{B}=0$.
Here, $\mathbf{J}$ is the electric current and $p$ is the pressure
of the plasma. Note that in the static configurations, 
the condition of magnetic confinement 
$\mathbf{E}=-\mathbf{v}\times\mathbf{B}$
just indicates the absence of the electric fields.

We introduce the Cartesian coordinates $(x, y, z)$
and assume the plane-symmetry in the $z$ direction.
Then, the magnetic field can be given by
%
\begin{equation}
\mathbf{B}=\mathbf{e}_z \times\nabla\Psi(x,y)+B_z(x,y)\mathbf{e}_z ,
\label{B-field}
\end{equation}
%
where $\Psi(x,y)$ is the so-called flux function and
$\mathbf{e}_z $ is the unit vector in the $z$ direction.
This formula satisfies the Gauss law automatically. By 
Ampere's law \eqref{Ampere}, the electric current is calculated as
%
\begin{equation}
\mathbf{J}=(\nabla^2\Psi)\mathbf{e}_z 
+(\nabla B_z)\times\mathbf{e}_z .
\label{J-flow}
\end{equation}
%
Substituting this formula into Eq.~\eqref{Force-balance}, we find
%
\begin{equation}
[(\nabla\Psi\times\nabla B_z)\cdot\mathbf{e}_z ]\mathbf{e}_z 
=\nabla p+(\nabla^2\Psi)\nabla\Psi+B_z\nabla B_z.
\end{equation}
%
Since the left hand side is the vector in the $z$ direction
while the right hand side is the vector in the $(x,y)$-plane,
both sides have to be zero:
%
\begin{equation}
(\nabla\Psi\times\nabla B_z)\cdot\mathbf{e}_z =0;
\label{EQ1}
\end{equation}
%
%
\begin{equation}
\nabla p+(\nabla^2\Psi)\nabla\Psi+B_z\nabla B_z=0.
\label{EQ2}
\end{equation}
%
Eq.~\eqref{EQ1} indicates $\nabla\Psi\parallel\nabla B_z$,
and then Eq.~\eqref{EQ2} indicates $\nabla\Psi\parallel\nabla p$.
Therefore, the contours of $B_z$, $p$, and $\Psi$
should coincide, and 
at least locally $B_z$ and $p$ are given by
%
\begin{equation}
B_z=f(\Psi),~~p=g(\Psi),
\label{def-fg}
\end{equation}
%
where $f$ and $g$ can be chosen arbitrarily as long as
they are regular and $g>0$.
Then, Eq.~\eqref{EQ2} is rewritten as
%
\begin{equation}
\nabla^2\Psi=-g^\prime-ff^\prime.
\label{GS1}
\end{equation}
%
This is the GS equation for the plane-symmetric case.

In this paper, we consider the situations where
the GS equation \eqref{GS1} is reduced to the following form:
%
\begin{equation}
\nabla^2\Psi=-\Psi\left(\alpha_0^2+\beta_0^2\Psi^2\right).
\label{GS2}
\end{equation}
%
Here, $\alpha_0$ and $\beta_0$ are assumed to be positive
without loss of generality. Since the source term has the cubic term,
we call this equation the cubic GS equation. 
Eq.~\eqref{GS2} can be derived if we choose the functions 
$f$ and $g$ satisfying the relation
%
\begin{equation}
f^2+2g=\alpha_0^2\Psi^2+\frac{1}{2}\beta_0^2\Psi^4
+C.
\label{relation-fg}
\end{equation}
%
Here, $C$ is a non-negative constant and it is zero 
if all physical quantities $\mathbf{B}$ and $p$ decay at the distant region. 
There are infinitely possible choices for $f$ and $g$,
since if $f=f_0$ and $g=g_0$ satisfy Eq.~\eqref{relation-fg},
$f=f_0+\chi$ and $g=g_0-f_0\chi-\chi^2/2$ also satisfy this relation,
where $\chi=\chi(\Psi)$ is an arbitrary function.
Therefore, one solution $\Psi$ of the GS equation \eqref{GS2}
can describe many different configurations.

It is possible to eliminate $\alpha_0$ and $\beta_0$
from Eq.~\eqref{GS2} by introducing the new coordinates
%
\begin{equation}
\bar{x}:=\alpha_0 x,~~\bar{y}:=\alpha_0 y
\label{xbarxybary}
\end{equation}
%
and the rescaled function 
%
\begin{equation}
u:=\frac{\beta_0}{\alpha_0}\Psi.
\label{u-Psi}
\end{equation}
%
By these transformations, the cubic GS equation becomes
%
\begin{equation}
u_{,\bar{x}\bar{x}}+u_{,\bar{y}\bar{y}}=-u(1+u^2).
\label{GS3}
\end{equation}
%
We require $u(\bar{x}, \bar{y})$ to be periodic 
in the $\bar{y}$ direction, 
to have the mirror symmetry about the $\bar{y}$
axis, and to become zero at $\bar{x}\to \infty$.
Namely, we look for the solutions which behave solitonlike 
in the $\bar{x}$ direction.

Let us study the asymptotic behavior of $u(\bar{x}, \bar{y})$ 
at $\bar{x} \to \infty$. Since we require that $u(\bar{x}, \bar{y})$
decay in this limit, Eq.~\eqref{GS3} is approximated as 
$u_{,\bar{x}\bar{x}}+u_{,\bar{y}\bar{y}}=-u$.
A solution to this equation satisfying the above requirements is
%
\begin{equation}
u=A\exp\left(-\bar{x}\sqrt{\hat{q}^{-2}-1}\right)\sin\left(\hat{q}^{-1}\bar{y}\right).
\label{asymptotic}
\end{equation}
%
Here, $A$ and $\hat{q}$ are constants and the value of $\hat{q}$
is limited as $0\le \hat{q}\le 1$. 
The period in the $\bar{y}$ direction is $\bar{y}_P=2\pi \hat{q}$.
This asymptotic behavior \eqref{asymptotic} satisfies $u(\bar{x},0)=0$
and $u_{,\bar{y}}(\bar{x},\bar{y}_P/4)=0$, and
we assume that these properties are held for all values of $\bar{x}$.
These two relations together with the condition
for the mirror symmetry $u_{,\bar{x}}(0,\bar{y})=0$
will become the boundary conditions in the numerical calculation.
Once the solution satisfying these boundary conditions 
is generated, the solution
in the whole region of $(\bar{x}, \bar{y})$
is obtained by the relation
$u(\bar{x},\bar{y})=-u(\bar{x},-\bar{y})=u(\bar{x},\bar{y}_P/4-\bar{y})
=u(\bar{x},\bar{y}+\bar{y}_P)=u(-\bar{x},\bar{y})$.

Note that Eq.~\eqref{asymptotic} is the exact solution
of the GS equation \eqref{GS2} in the case $\alpha_0=1$
and $\beta_0=0$.
Therefore, without the cubic term in Eq.~\eqref{GS3}, 
the solution diverges at $\bar{x}\to -\infty$.
However, in the case where the cubic term is present, the solution
$u$ having the mirror symmetry about the $\bar{y}$ axis
can exist because of the nonlinear effect.
It will be explicitly shown in Sec. IV.

%
%
\section{Numerical calculation}

In this section, we explain how to calculate 
the solitonlike solution $u$ of the cubic GS equation~\eqref{GS3}. 
The numerical method is explained in Sec. IIIA. 
In order to establish the existence of the solitonlike solutions, 
we have to check the numerical errors carefully. This is discussed
in Sec. IIIB. 

%
%
\subsection{Numerical method}

In the numerical calculation, it is very convenient to 
choose the coordinates $(X, Y)$ that are normalized
by a quarter of the period in the $\bar{y}$ direction.
For this reason, we introduce a parameter
%
\begin{equation}
q:=({\pi}/{2})\hat{q}
\end{equation} 
%
and perform the coordinate transformation
%
\begin{equation}
X:=q^{-1}\bar{x},~~Y:=q^{-1}\bar{y}.
\label{XxbarYybar}
\end{equation}
%
In the coordinates $(X, Y)$, Eq.~\eqref{GS3} becomes
%
\begin{equation}
u_{,XX}+u_{,YY}=-q^2u(1+u^2),
\label{GS4}
\end{equation}
%
and the period in the $Y$ direction is $Y_P=4$.
By the symmetries of the solution that we required in Sec. II,
it is sufficient to solve in the range $0\le X\le X_{\rm max}$
and $0\le Y\le 1$. Here, $X=X_{\rm max}$ is the outer boundary of
the region of numerical calculation, and we choose
$X_{\rm max}=5$. The error coming from this cutoff value 
will be estimated in the next subsection.

The boundary conditions are  
$u=0$ at $Y=0$, $u_{,Y}=0$ at $Y=1$, and $u_{,X}=0$ at $X=0$.
At the outer boundary $X=X_{\rm max}$, we have to impose the condition
\eqref{asymptotic}, which is 
$u=A\exp\left(-X\sqrt{\pi^2/4-q^2}\right)\sin(\pi/2)Y$ 
in the $(X, Y)$ coordinates. 
Because we do not know the value of $A$
before generating the solution of $u$, we calculate $u_{,X}$
and eliminate $A$. This leads to the so-called Robin boundary condition
%
\begin{equation}
u_{,X}=-u\sqrt{\pi^2/4-q^2}.
\label{BC}
\end{equation}
%
This formula is used as the boundary condition at $X=X_{\rm max}$.

In order to solve Eq.~\eqref{GS4} numerically,
we adopted the second-order finite difference scheme with uniform grids.
Since Eq.~\eqref{GS4} is an elliptic equation, 
we have to prepare an initial surface and 
make it converge to the solution by the method of iteration. 
However, the solution was found to be unstable against this process.
This is in contrast to the case of Ref.~\cite{YSS06}, 
where one of us solved a similar equation with no problem. 
The reason for the difference between the two cases is as follows. 
In both cases, the equation has the form $\nabla^2u=-F(u)$.
The function $F(u)$ is monotonically decreasing as $u$ grows
in the case of \cite{YSS06}, while it is a monotonically increasing function
in the present case as found from Eq.~\eqref{GS4}.
Let us consider what happens in the latter case. 
Suppose the initial surface $u_0$
is slightly larger than the real solution $\hat{u}$. The program
makes the surface approach the solution of the equation 
$\nabla^2u=-F(u_0)$. Because $F(u_0)>F(\hat{u})$,
the solution of this equation $u_1$ is further larger than 
$u_0$, i.e. $u_1>u_0>\hat{u}$. Therefore, by continuing these processes, 
the value of $u$ becomes larger and larger and eventually diverges. 
On the other hand, if $u_0$ is a little smaller than $\hat{u}$, the value of
$u$ becomes smaller and smaller and 
collapses to $u=0$, i.e. the trivial solution. 
For this reason, the nontrivial solution of Eq.~\eqref{GS4} is numerically 
unstable, and a new idea is needed to obtain it.

Although we could not develop a new code which can
automatically generate such numerically unstable solutions, we 
found a procedure to realize the sufficient convergence
of iterations.
The point is that for some initial surface $u_0$, the 
surface $u$ approaches the real solution $\hat{u}$ to some
extent and then leaves it after that in the process of iteration. 
In other words, the real solution $\hat{u}$ behaves like 
an intermediate attractor in this process. 
Such a behavior typically occurs when $u_0$ crosses
the real solution $\hat{u}$, i.e. the regions $u_0>\hat{u}$ 
and $u_0<\hat{u}$ both exist. Using this property,
we proceeded as follows. We prepare a good initial surface $u^{(0)}$
and start the computation (the first trial). While the program is running,
we observe the convergence parameters
%
\begin{equation}
\Delta_{1}:=\frac{\sum |\Delta u_{(I,J)}|}{\sum|u_{(I,J)}|},
~~
\Delta_{2}:=\max\left|\frac{\Delta u_{(I,J)}}{u_{(I,J)}}\right|,
\end{equation}
%
where $(I,J)$ are the label of the grids and $\Delta u_{(I,J)}$
denotes the difference from the finite difference equation.
The values of $\Delta_{1,2}$ first decrease and then increase. 
Just before $\Delta_1$ starts to increase, we write down the surface
$u^{(0)}_{F}$ and stop the program.
Then, we prepare the new initial surface $u^{(1)}$ by 
$u^{(1)}=(1+\epsilon)u^{(0)}_{F}$ and run the program again 
(i.e. the second trial). Choosing $\epsilon$ properly,
we can make $\Delta_{1,2}$ further smaller although a little
experience is required in order to find the effective value of $\epsilon$.
We continued these processes of trials until the conditions
$\Delta_1<10^{-8}$ and $\Delta_2<10^{-7}$ are achieved.

%
%
\subsection{Error estimates}

Using the above technique, we solved the cubic GS equation~\eqref{GS4}
for $\hat{q}=0.1$--$0.9$ with $0.1$ intervals.
In all cases, we adopted the grid numbers $(250\times 50)$.
Since the solution is numerically unstable, we have to 
check the numerical error carefully in order to prove
that our solution is not a numerical artifact. 
There are three sources of the numerical errors: 
the finiteness of $X_{\rm max}$, the finiteness of the
grid sizes, and the truncation of the convergence process.

The error by the finite $X_{\rm max}$ value
is evaluated by $\delta_1=[u(X_{\rm max}, 1)]^2$, since
the boundary condition \eqref{asymptotic} is derived by
ignoring the cubic term in Eq.~\eqref{GS3}. 
The value of $\delta_1$ is less than $ 0.02\%$ for $0.1\le\hat{q}\le 0.7$.
It becomes larger as $\hat{q}$ is increased and $0.3\%$
for $\hat{q}=0.9$. This is because the value of $u$ decays very slowly
for $\hat{q}\simeq 1$ by the boundary condition \eqref{asymptotic}.
We also compared the results of $X_{\rm max}=5$
and $10$ in the case $\hat{q}=0.9$.
The error estimated in this way is $0.1\%$.

The error by the finite grid sizes is estimated as 
$0.03\%$ for all values of $\hat{q}$
by comparing the results of $(250\times 50)$ and
$(125\times 25)$ grid numbers. 
This is natural because we used the
second-order accuracy scheme and thus the error
is expected to have the order of the squared grid size $\sim 0.04\%$.

The error by the truncation of the convergence process
is estimated as follows. Suppose $u_{F}$ is the obtained
solution and consider the equation
$\nabla^2u=-q^2u_F(1+u_F^2)$. 
If the convergence is perfect,
the solution of this equation is $u=u_F$.
Since the convergence process is truncated 
by the criterion explained above, 
the actual solution of $u$ is different from $u_F$,
and this difference indicates the error amount. 
In this way, the error is estimated to be less than 
$0.03\%$ for all values of $\hat{q}$.

Therefore, all the numerical errors are small and our 
results are reliable.

%
%
\section{Numerical results}

\begin{figure}[tb]
\centering
{
\includegraphics[width=0.45\textwidth]{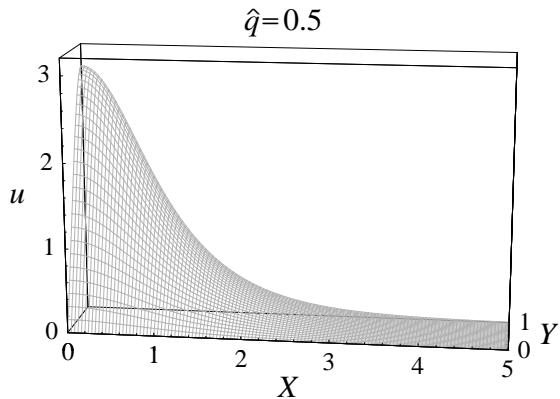}
}
\caption{3D plot of the generated solution $u$ for $\hat{q}=0.5$ 
in the region $0\le X\le 5$ and $0\le Y\le 1$.}
\label{3d-amp-L0.5}
\end{figure}

\begin{figure}[tb]
\centering
{
\includegraphics[width=0.45\textwidth]{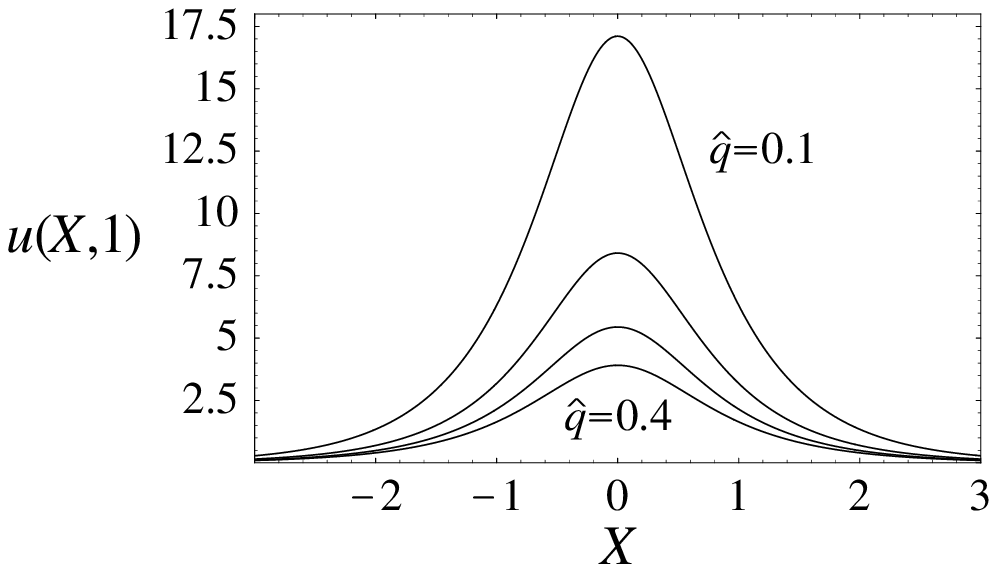}
\includegraphics[width=0.45\textwidth]{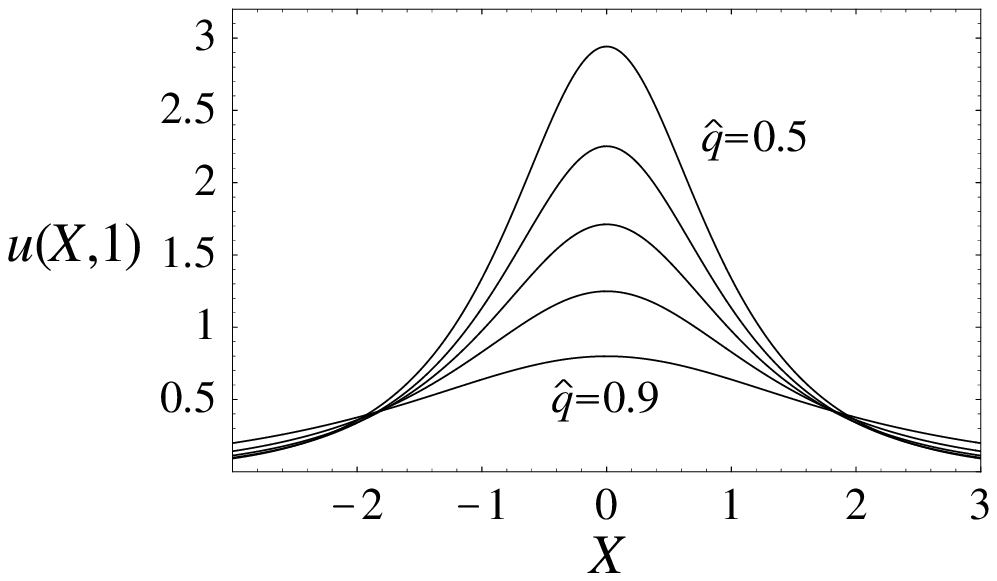}
}
\caption{The behavior of $u(X,1)$ on the line $Y=1$ 
for $\hat{q}=0.1$--$0.4$ (left) and $\hat{q}=0.5$--$0.9$ (right).}
\label{sec-Y1.0}
\end{figure}

\begin{figure}[tb]
\centering
{
\includegraphics[width=0.45\textwidth]{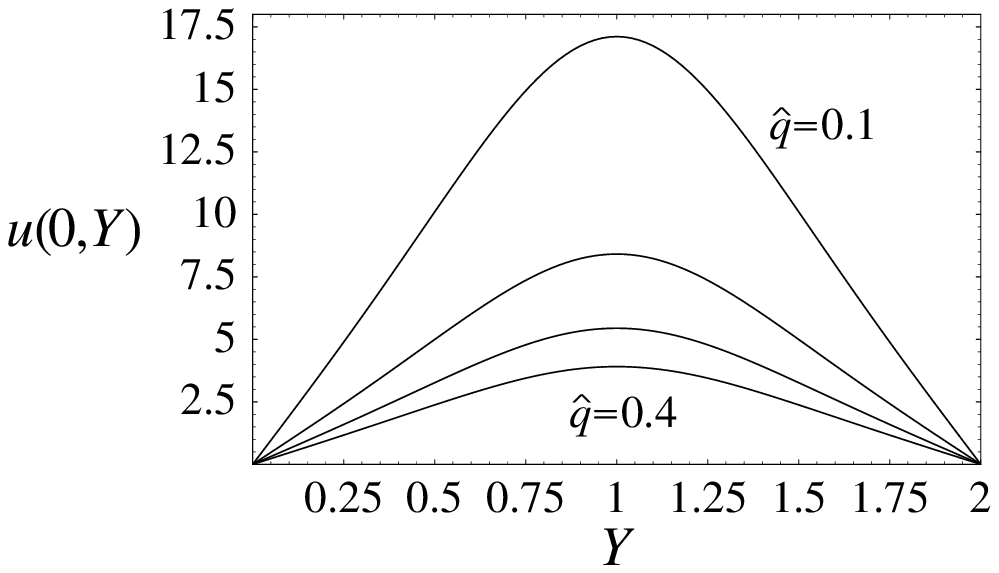}
\includegraphics[width=0.45\textwidth]{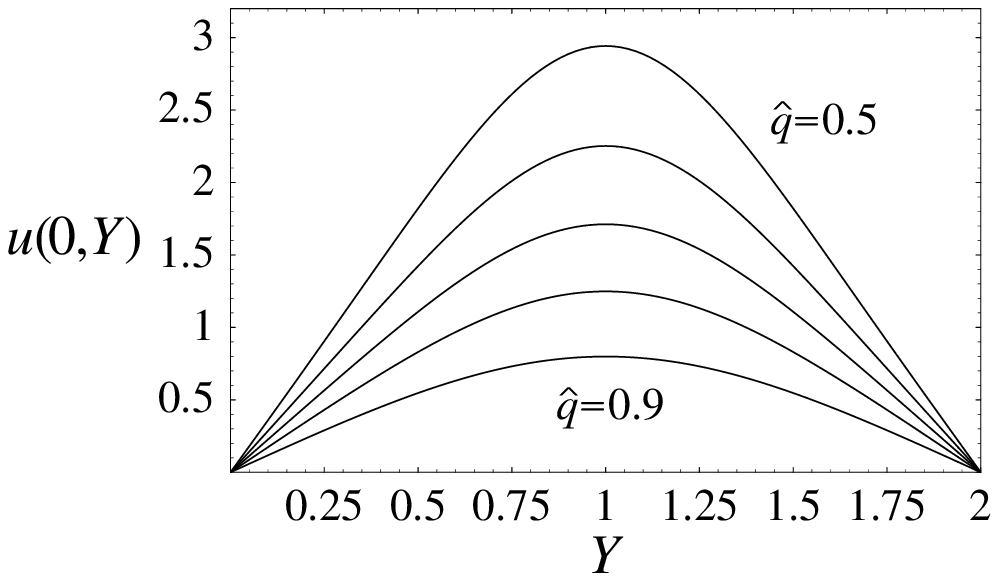}
}
\caption{The behavior of $u(0,Y)$ on the $Y$-axis 
for $\hat{q}=0.1$--$0.4$ (left) and $\hat{q}=0.5$--$0.9$ (right).}
\label{sec-X0.0}
\end{figure}

Now we show the numerical results. Figure~\ref{3d-amp-L0.5}
shows the 3D plot of the numerical solution of $u$
in the range $0\le X\le 5$ and $0\le Y\le 1$ for $\hat{q}=0.5$.
The solution $u(X,Y)$ takes its maximum value $u_{peak}$ at $(X,Y)=(0,1)$.
Figure~\ref{sec-Y1.0} shows the behavior
of $u(X,1)$ on the line $Y=1$ and Fig.~\ref{sec-X0.0}
shows the behavior of $u(0,Y)$ on the line $X=0$ (i.e. $Y$-axis)
for $\hat{q}=0.1$--$0.9$. The peak value $u_{peak}$
increases as $\hat{q}$ is decreased.
This is because the right hand side of Eq.~\eqref{GS4}
is proportional to $q^2$ and thus larger value of $u$
is necessary for smaller $q$
in order that the effect of the nonlinear term becomes relevant. 
From the right plot of Fig.~\ref{sec-Y1.0}, we see that
the value of $u(X,1)$ decays more slowly for larger $\hat{q}$
because of the boundary condition \eqref{BC}.

%
%

%
%

%
%

\begin{figure}[tb]
\centering
{
\includegraphics[width=0.45\textwidth]{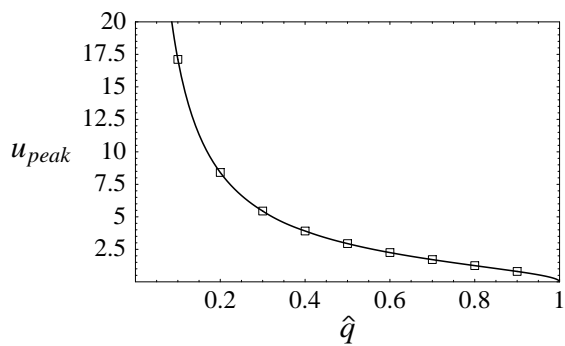}
}
\caption{The dependence of the peak value $u_{peak}$ on $\hat{q}$. 
The numerical data is shown by squares ($\square$). $u_{peak}$ behaves
as $u_{peak}\propto \hat{q}^{-1}$ and $\sqrt{1-\hat{q}}$  
in the neighborhood of $\hat{q}=0$ and $1$, respectively.}
\label{peak-amplitude}
\end{figure}

Figure~\ref{peak-amplitude} shows the dependence of the peak value
$u_{peak}$ on $\hat{q}$. From this figure, it is understood that $u_{peak}$
diverges in the limit $\hat{q}\to 0$. By plotting 
the relation between $\hat{q}u_{peak}$ and $\hat{q}$, we found that
$u_{peak}$ is approximated by $u_{peak}\simeq 1.72/\hat{q}$
for small $\hat{q}$. 
The solution of $u$ becomes $u\equiv 0$ in the limit $\hat{q} \to 1$, because  
$u$ depends only on $Y$ in this limit by the boundary condition \eqref{BC}
while we are solving the sequence for which $u=0$ at $X=\infty$. 
By plotting the values of $u_{peak}^2$ as a function of $1-\hat{q}$,
we found that the formula
$u_{peak}\simeq 2.27 \sqrt{1-\hat{q}}$ 
approximately holds in the neighborhood of $\hat{q}=1$.

%
%

We summarize the general properties that do not depend
on specific forms of $f$ and $g$.
From Eqs.~\eqref{xbarxybary} and \eqref{XxbarYybar}, 
the coordinates $(X,Y)$ and $(x,y)$ are related
as $X=\alpha_0q^{-1}x$ and $Y=\alpha_0q^{-1}y$.
In the $(x,y)$ coordinates, the period of $\Psi$ in the $y$ direction is
%
\begin{equation}
y_P=\frac{4q}{\alpha_0}.
\label{yP}
\end{equation}
%
By the dimensional analysis, 
$f(\Psi)$ and $g(\Psi)$ are found to be expressed as 
%
\begin{equation}
f(\Psi)=\frac{\alpha_0^2}{\beta_0}\hat{f}(u),~~
g(\Psi)=\frac{\alpha_0^4}{\beta_0^2}\hat{g}(u).
\end{equation}
%
Here, by Eq.~\eqref{relation-fg}, $\hat{f}(u)$ and $\hat{g}(u)$
are related as
%
\begin{equation}
\hat{f}^2+2\hat{g}=u^2+\frac{u^4}{2}+\hat{C},
\end{equation}
%
where $\hat{C}$ is a non-negative constant. 
Calculating the magnetic field \eqref{B-field} 
and the electric current \eqref{J-flow}
using Eqs.~\eqref{def-fg}, \eqref{u-Psi} and \eqref{GS2},
we obtain
%
\begin{equation}
\mathbf{B}=\frac{\alpha_0^2}{\beta_0}
\left[
\frac{1}{q}
\left(
- u_{,Y}\mathbf{e}_x + u_{,X}\mathbf{e}_y
\right)
+\hat{f}(u)\mathbf{e}_z
\right],
\label{B-expression}
\end{equation}
%
%
\begin{equation}
\mathbf{J}=-\frac{\alpha_0^3}{\beta_0}
\left[\frac{\hat{f}_{,u}}{q}
\left(
- u_{,Y}\mathbf{e}_x + u_{,X}\mathbf{e}_y
\right)
+u(1+u^2)\mathbf{e}_z
\right].
\label{J-expression}
\end{equation}
%
From these formulas, the meanings of 
the parameters $q$, $\alpha_0$, and $\beta_0$ are understood.
Since the inside of the parenthesis of 
Eq.~\eqref{B-expression} depends only on $u$ and $q$ 
(for a fixed form of $\hat{f}$), the direction of 
the magnetic field at a given position $(X,Y)$ is determined 
once the value of $q$ is specified. 
This means that the shape similarity of the field lines is 
preserved when  $\alpha_0$ and $\beta_0$ are varied.
Furthermore, nondimensional quantities such as 
the beta ratio $2p/B^2$ are independent of $\alpha_0$ and $\beta_0$. 
Therefore, $q$ is the parameter
that determines all nondimensional properties
of the system. For a fixed $q$, the value of $\alpha_0$
determines the characteristic scale of the system 
through Eq.~\eqref{yP}. 
After fixing $q$ and $\alpha_0$,
the magnitude of $\mathbf{B}$ is determined
by specifying $\beta_0$. 
Hence $\beta_0$ is (say) the field strength parameter.

%
\begin{figure}[tb]
\centering
{
\includegraphics[width=0.45\textwidth]{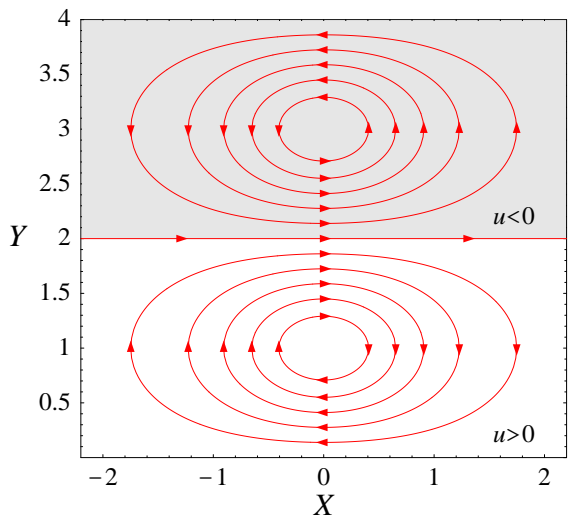}
}
\caption{The contours of $|u|=0.0$--$2.5$
(with $0.5$ intervals)  for $\hat{q}=0.5$ on the $(X,Y)$-plane.
The arrows indicate the directions of magnetic field lines. 
They are clockwise in the region $u>0$
and counter-clockwise in the region $u<0$.}
\label{flowB}
\end{figure}
%

From $x$ and $y$ components of Eqs.~\eqref{B-expression}
and \eqref{J-expression}, the magnetic field lines
and the electric currents are confined on the contour
surfaces of $u$. 
Figure~\ref{flowB} shows the contours of $u$
on the $(X,Y)$-plane for $\hat{q}=0.5$. 
The directions of the magnetic fields are also shown.
The magnetic fields are clockwise in the region $u>0$
and counter-clockwise in the region $u<0$.
From Eq.~\eqref{J-expression}, it is seen that
$J_z<0$ in the region $u>0$ and 
$J_z>0$ in the region $u<0$.
This relation between the directions of 
$(B_x, B_y)$ and 
the sign of $J_z$ is consistent with Ampere's law.

The $z$ component of the magnetic field 
is specified by the function $\hat{f}(u)$. 
Changing $\hat{f}(u)$ affects $(J_x, J_y)$ through Ampere's law.
If $\hat{f}(0)=0$ and $\hat{f}(u)$ is a monotonic function 
in each region of $u>0$ and $u<0$,
a simple relation exists between the sign of $B_z$ 
and the directions of $(J_x, J_y)$.
Let us consider the region $u>0$. If $\hat{f}(u)$ is a monotonically 
increasing function, we have $\hat{f}>0$ and $\hat{f}_{,u}>0$.
This indicates that $B_z>0$ and the electric currents
are counter-clockwise. On the other hand, if $\hat{f}(u)$ is a monotonically 
decreasing function, we have $\hat{f}<0$ and $\hat{f}_{,u}<0$,
which means that $B_z<0$ and the electric currents
are clockwise. The same relation 
is obtained also for $u<0$.

%
%
\section{Summary and discussion}

In this paper, we studied the solitonlike solutions of
magnetostatic equilibria by numerically solving  
the cubic GS equation. Although the solutions were 
unstable against the numerical iteration, we found 
the procedure to realize the sufficient convergence
and obtained the highly accurate solutions. 
The generated solutions are solitonlike in the $x$ direction,
periodic in the $y$ direction and symmetric in the $z$ direction.
Our result proves the existence of the solitonlike solution
that was questioned in recent years \cite{Lap03, THT04, Lap04, SSP05, LK05}.

The solitonlike solution obtained in this paper behaves as
an even function on a $y=\mathrm{const.}$ line
and has one extreme at the center.
It is interesting to examine the existence of
another solution that behaves as an odd function
on a $y=\mathrm{const.}$ line and has an extreme
in each region of $x>0$ and $x<0$.
Such a (say) $2$-solitonlike solution could be expected 
by the following discussion. Denoting the obtained solitonlike solution
by $\Psi^{(1)}(x,y)$, the function
%
\begin{eqnarray}
\Psi^{(2)}(x,y)&=&\Psi^{(1)}(x-d/2, y)-\Psi^{(1)}(x+d/2,y)
\end{eqnarray}
%
also approximately satisfies 
the GS equation~\eqref{GS2} for sufficiently large $d$, 
since $\Psi^{(1)}(x)$ decays exponentially for large $|x|$. 
Therefore, one might expect 
the existence of $2$-solitonlike solutions also for finite values of $d$.
However, it is possible to show that  
no 2-solitonlike solution exists 
under the boundary condition
$\Psi(0,y)=0$ for any $f$ and $g$ satisfying $f(0)f^\prime(0)+g^\prime(0)=0$.
To show this, we multiply $\Psi_{,x}$ to the GS equation \eqref{GS1} as
%
\begin{equation}
\Psi_{,xx}\Psi_{,x}+\Psi_{,yy}\Psi_{,x}
=-(ff^\prime+g^\prime)\Psi_{,x},
\end{equation}
%
and integrate this equation over the region $0\le x\le \infty$
and $0\le y\le y_P$.  Assuming the exponential decay
of $\Psi$ at $x\gg 1$, the integrals of the second term on the left hand side
and the right hand side vanish, and we have
%
\begin{equation}
\int_0^{y_P} \Psi_{,x}^2(0,y) dy=0,
\end{equation}
%
and therefore $\Psi_{,x}(0,y) = 0$. 
Then, the GS equation \eqref{GS1}
indicates that all derivatives of $\Psi$ with respect to $x$
vanish on the symmetry axis (assuming $\Psi$ to be analytic). 
Hence $\Psi= 0$ is the only solution.
Physically, this means that when two or more solitons coexist,
they interact each other and cannot be in equilibrium.

It is interesting to discuss the stability of the solitonlike solution 
obtained in this paper.
The system is expected to be unstable, since
the magnetic islands are periodically located in the $y$ direction
and interactions between them are present.
The most important factor for such interactions
is the directions of the electric currents of the islands.
If the currents of the islands are parallel (i.e., $J_z$ has the same sign),
the islands attract each other and  coalesce 
into larger islands. Such instability 
is known as the coalescence instability \cite{FK77,PW79}.  
On the other hand, if the currents of the neighboring islands
are anti-parallel (i.e., $J_z$ has an alternating sign),
their interaction is repulsive and they tend to repel each other in the $x$ direction
as a result of small disturbance.
Since $J_z$ has an alternating sign in our system
as seen from Eq.~\eqref{J-expression}, the
repulsive instability is expected.
In fact, both instabilities were confirmed by the recent 
numerical work on the dynamics of magnetic islands
with parallel and anti-parallel currents \cite{LK05}.

Although the plane-symmetric solitonlike solution
in this paper could be of use in the contexts of the astrophysics
or the solar physics, it would be more interesting
to apply our method to the axisymmetric case.
In the observations of active galactic nuclei (AGN), the astrophysical jets
are often found to have knotty structures
that suggest the presence of magnetic multiple islands \cite{BHLBL94}.
Several models of the knotty jets have been proposed, and 
one of the possible directions is to model
the knotty jets as magnetostatic equilibria \cite{KC85,B00,Lap03, Lap04,LK05}.
Although these studies do not give the mechanism for the formation of
the knotty jets, such models are expected to  
explain the long lifetime of collimation and knotty structure
simultaneously. Namely, the knotty jets can maintain
their shapes because they are in equilibrium
in the comoving frame, and the time scale of the
instabilities gives the lifetime of the knotty structure.
The authors of \cite{LK05} studied the growth of instabilities
of plane-symmetric solitonlike configurations
by performing numerical simulations.
Assuming that the plane-symmetric solitonlike systems
well approximate the axisymmetric ones,
they compared the results with the observations of the knotty jet
of the radio galaxy 3C 303 \cite{K76, LP91, KEGTW03}. 
Their conclusion is 
that the numerical simulation gives good agreement with
the actual observations. Here,
it should be pointed out that the assumption in that paper
is not obvious and has to be justified. 
For this reason, the extension to the axisymmetric cases
is necessary in order to examine if the solitonlike solutions
can really model the astrophysical knotty jets.
An axisymmetric quasiperiodic magnetostatic solution
of the linear GS equation was proposed as the
astrophysical jet model \cite{B00}.
The numerical method in this paper enables us
to generalize the study of \cite{B00} to the case 
of the nonlinear axisymmetric GS equation and thus to obtain
further large class of astrophysical jet models
as magnetostatic equilibria. 
The present work is the first step toward this direction, and 
we are planning to generalize our result to the axisymmetric case.
It would be also interesting
to further explore the solitonlike solutions in the helically symmetric cases.

\acknowledgments

HY thanks the Killam Trust for financial support.



\end{document}